\pdfoutput=1
\documentclass[a4paper,11pt]{article}
\usepackage{amsmath,amssymb,amsfonts}
\usepackage[multiple,stable]{footmisc}
\usepackage[amsmath,amsthm,thmmarks]{ntheorem}
\allowdisplaybreaks[4]
\usepackage[noconfig]{refstyle}
\usepackage[dvipsnames]{xcolor}
\usepackage[hyperfootnotes=false, linktocpage=true, colorlinks, citecolor=blue, linkcolor=blue, urlcolor=Maroon]{hyperref}
\usepackage{array,arydshln,tabularx,booktabs,geometry,subfigure,graphicx,longtable}
\usepackage[bf]{caption}
\usepackage{tikz}
\usepackage{cite}

\usetikzlibrary{shapes,arrows}
\tikzstyle{block} = [rectangle, draw, text width=7em, text centered, rounded corners, minimum height=3em]

\usepackage{cite}
\usepackage{arydshln}
\usepackage{tikz}
\usetikzlibrary{positioning,automata,decorations.pathreplacing,shapes}
\usepackage[english]{babel}
\usepackage{microtype}

\usepackage[all,cmtip]{xy}
\usetikzlibrary{matrix, arrows}

\let\eqref=\relax
\newref{eq}{name={eq.~},Name={Eq.~},names={eqs.~},Names={Eqs.~},rngtxt={-},refcmd=(\ref{#1})}
\newref{f}{name={footnote~},Name={Footnote~},names={footnotes~},Names={Footnotes~}}
\newref{app}{name={appendix~},Name={Appendix~},names={appendixes~},Names={Appendixes~}}
\newref{tab}{name={table~},Name={Table~},names={tables~},Names={Tables~}}
\newref{ch}{name={chapter~},Name={Chapter~},names={chapters~},Names={Chapters~}}
\newref{sec}{name={section~},Name={Section~},names={sections~},Names={Sections~}}
\newref{fig}{name={figure~},Name={Figure~},names={figures~},Names={Figures~}}
\numberwithin{equation}{section}

\newcommand{\be}{\begin{equation}}
\newcommand{\ee}{\end{equation}}
\newcommand{\bea}{\begin{equation}\begin{aligned}}	
\newcommand{\eea}{\end{aligned}\end{equation}}		

\newcommand{\iddots}{\mathinner{\mkern2mu\raise1pt\hbox{.}\mkern2mu \raise4pt\hbox{.}\mkern2mu\raise7pt\hbox{.}\mkern1mu}}

\providecommand{\id}{\leavevmode\hbox{\small$\mathrm{1}$\kern-3.8pt\normalsize$\mathrm{1}$}}
\def\fnote#1#2{\begingroup\def\thefootnote{#1}\footnote{#2}
     \addtocounter{footnote}{-1}\endgroup}

   \newcommand{\eref}[1]{(\ref{#1})}
   \newcommand{\eeq}{\end{equation}}
\newcommand{\beq}{\begin{equation}}
\newcommand{\ba}{\begin{array}}
\newcommand{\ea}{\end{array}}

\newcommand{\nn}{\nonumber}

\begin{document}

\vspace{1cm}

\title{       {\Large \bf Fibrations in Non-simply Connected Calabi-Yau Quotients}}

\vspace{2cm}

\author{
Lara~B.~Anderson,${}{}$
James~Gray,${}{}$ and
Brian~Hammack${}{}$
}
\date{}
\maketitle
\begin{center} {\small ${}${\it Department of Physics, 
Robeson Hall, Virginia Tech \\ Blacksburg, VA 24061, U.S.A.}}

\fnote{}{lara.anderson@vt.edu }
\fnote{}{jamesgray@vt.edu}
\fnote{}{bh@vt.edu}

\end{center}

\begin{abstract}
\noindent
In this work we study genus one fibrations in Calabi-Yau three-folds with a non-trivial first fundamental group. The manifolds under consideration are constructed as smooth quotients of complete intersection Calabi-Yau three-folds (CICYs) by a freely acting, discrete automorphism. By probing the compatibility of symmetries with genus one fibrations (that is, discrete group actions which preserve a local decomposition of the manifold into fiber and base) we find fibrations that are inherited from fibrations on the covering spaces. Of the 7,890 CICY three-folds, 195 exhibit known discrete symmetries, leading to a total of 1,695 quotient manifolds. By scanning over 20,700 fiber/symmetry pairs on the covering spaces we find 17,161 fibrations on the quotient Calabi-Yau manifolds. It is found that the vast majority of the non-simply connected manifolds studied exhibit multiple different genus one fibrations - echoing a similar ubiquity of such structures that has been observed in other data sets. The results are available at \cite{quotientlist}. The possible base manifolds are all singular and are catalogued. These Calabi-Yau fibrations generically exhibit multiple fibers and are of interest in F-theory as backgrounds leading to theories with superconformal loci and discretely charged matter.
\end{abstract}

\thispagestyle{empty}
\setcounter{page}{0}
\newpage

\tableofcontents

\section{Introduction}

The study of Calabi-Yau (CY) manifolds has long been a central topic in the study of string compactifications. In recent years, the fibration structures exhibited by these manifolds has been extensively studied for a number of reasons. First, in F-theory genus one fibrations are essential to capture the  ``geometrized" axio-dilaton of Type IIB string theory \cite{Vafa:1996xn}, describing this physical degree of freedom as the complex structure of a Torus fiber of a Calabi-Yau manifold. The structure of the low energy effective physics seen in F-theory compactifications is therefore crucially linked to the nature of the genus one fibrations of Calabi-Yau spaces. Second, fibrations of various dimensionalities in Calabi-Yau manifolds play a crucial role in the subject of string dualities. They are central in heterotic/Type IIA duality, heterotic/F-theory duality and F-/M-theory duality, and many others.   

Mathematically, genus one fibered geometries are also important because they supply a foothold into attempts to \emph{classify all Calabi-Yau manifolds} in fixed dimension. The set of genus one fibered Calabi-Yau three-folds is proven to be finite \cite{gross_finite}. More recent work \cite{2016arXiv160802997D} has investigated similar results for genus one fibered Calabi-Yau four-folds and five-folds. A key motivation for these classifications \cite{grassi,gross_finite} is the hope that they could be used as a first step in establishing the finiteness of all Calabi-Yau manifolds of a given dimension. Despite this, and despite the clear utility of fibered Calabi-Yau manifolds in the subject of string dualities, it was historically thought that fibration structures (which include genus-one, $K3$ or abelian surface fibrations), would likely be rare within the set of all such geometries.

Recent studies have shown that, in fact, nearly all known Calabi-Yau manifolds are genus-one fibered \cite{Rohsiepe:2005qg,Johnson:2014xpa,Gray:2014fla,Johnson:2016qar,Candelas:2012uu,Anderson:2016cdu,Anderson:2017aux}. Indeed, generically, it appears that all Calabi-Yau manifolds which have been constructed using standard methods can be written as genus-one fibrations, generically in multiple, inequivalent ways, over potentially topologically distinct base manifolds \cite{Rohsiepe:2005qg,Gray:2014fla,Anderson:2016cdu,Morrison:2016lix}. A Calabi-Yau manifold can be called \emph{multiply elliptically fibered} (or genus one fibered in the case where the fibration does not admit a section\footnote{See \cite{Morrison:2014era,Anderson:2014yva} for some of the physical distinctions between compactifications on elliptically or genus one fibered three-folds.}) if it admits multiple descriptions of the form $\pi_i: X_{n} \longrightarrow B^{(i)}_{n-1}$ where the generic fiber is elliptic (or genus-one) $\mathbb{E}_{(i)b}=\pi^{-1}(b\in B^{(i)}_{n-1})$ (denoted succinctly by $\pi_i: X_{n} \stackrel{\mathbb{E}_{(i)}}{\longrightarrow} B^{(i)}_{n-1}$). In other words, we can draw a diagram such as the following.
\begin{equation} \label{manyfibs}
\xymatrix{
& X_{n} \ar[ld]_{\mathbb{E}_{(1)}} \ar[d]^{\mathbb{E}_{(2)}}  \ar[rd]^{\mathbb{E}_{(i)}} &\\
B^{(1)}_{n-1} & B^{(2)}_{n-1} \ldots & B^{(i)}_{n-1}}
\end{equation}

For each different fibration, $\pi_i$, the associated Weierstrass model \cite{nakayama}, the structure of the singular fibers, the discriminant locus, fibral divisors, the topology of the base manifolds $B^{(i)}_{n-1}$ and the Mordell-Weil group can all be different \cite{Anderson:2016cdu,Anderson:2017aux}. These prolific and ubiquitous multiple fibration structures were also studied for CICY four-folds in \cite{Gray:2014fla,Lynker:1998pb,Brunner:1996bu,cicylist4}. 

In the present work, it is our aim to extend the results above by studying fibrations in a different class of Calabi-Yau geometries. Although it is true that the vast majority of manifolds that have been studied have displayed a rich and ubiquitous multiple fibration structure, all of these manifolds have been constructed in a similar manner and it is possible that they suffer from a ``lamp post" effect. One common property of the fibered manifolds studied to date has been that almost all are simply connected\footnote{Within the Kreuzer-Skarke database all but 16 of the hypersurfaces in toric varieties are simply connected \cite{Kreuzer:2000xy}.}. Given this, it is of interest to ask whether the results about CY fibrations that have been obtained in the literature thus far carry over to other constructions and, in particular, to the case of non-simply connected manifolds?

The simplest way in which to construct a large class of non-simply connected Calabi-Yau three-folds is to take an existing simply connected dataset of manifolds and quotient them by freely acting discrete symmetries, $\Gamma$ (note that such quotients are not Calabi-Yau manifolds in even dimensions and thus we can not follow such a route in those cases). If $|\Gamma|=n$, the relationship between the ``upstairs" manifold $X$ (simply connected) and its ``downstairs" quotient, $\hat{X}$ is that of an $n$-sheeted covering:
\beq\label{quotientdef}
q: X \to \hat{X}=X/\Gamma~.
\eeq
Thus, in this work we will focus on studying a particular dataset of genus one fibered Calabi-Yau three-folds - \emph{non-simply connected manifolds that are constructed as smooth quotients of Calabi-Yau three-folds by a freely acting discrete automorphism}. To this end, we have employed the only dataset of three-folds for which a large class of discrete automorphisms has been classified \cite{Braun:2010vc} -- the Calabi-Yau three-folds defined as complete intersections in products of ordinary projective spaces \cite{Candelas:1987kf}. 

The genus one fibration structures of these geometries is virtually unexplored. In addition, the physical properties of effective theories obtained by F-theory compactifications on these spaces has only recently begun to be studied \cite{Anderson:2018heq}. In these geometries, although the base of the fibration is generically singular, the total space is smooth. This smooth total space is possible due to the presence of multiple fibers over the singular points on the base. The combination of singular base manifolds to the fibration and isolated multiple fibers leads to interesting physics: namely the presence of superconformal sectors in the lower dimensional effective field theory, coupled to discretely charged matter. It should be noted that if the discrete symmetry, $\Gamma$, being employed in the quotient \eref{quotientdef} does not act in a simple toric fashion, then the quotient manifolds being constructed are not themselves obviously describable in terms of a hypersurface or complete intersection in a toric variety, and thus the manifolds constructed are quite distinct from the usual complete intersection datasets and could provide novel features.

In studying these geometries, we aim to achieve two goals. First, we would like to answer the question \emph{does the trend of ubiquitous genus one fibrations seen in other constructions of Calabi-Yau manifolds continue in this non-simply connected data set?} Moreover, are virtually all of the manifolds multiply genus one fibered as seen in other cases? Second, we hope to provide a large class of examples for researchers who wish to study the physics associated to F-theory compactifications on non-simply connected total spaces with multiple fibers \cite{Anderson:2018heq}.

Our results include the following:
\begin{itemize}
\item We study the fibrations of 1,695 non-simply connected Calabi-Yau three-folds. A scan for fibrations (descending from the covering spaces) yields 17,161 fibrations for the dataset. As with other datasets of Calabi-Yau three-folds, we find that once again nearly all manifolds (more than $95\%$ in this dataset) appear to admit at least one (and generically multiple) descriptions as genus one fibrations.
\item With a view towards F-theory compactifications, we characterize the maximal discrete symmetry orders that are compatible with quotienting and fibration structures in this dataset, as well as the possible base manifolds that can occur.
\end{itemize}

The rest of this paper is structured as follows. In the next section, we review the CICY construction, discrete symmetries of CICYs, fibrations in CICYs, and how it can be determined whether fibration and symmetry structures are compatible, leading to non-simply connected manifolds in \eref{quotientdef} that are genus one fibered. In Section \ref{res1}, we provide the results of an exhaustive study of the obvious fibration structures associated to quotients of CICY three-folds. In Section \ref{res2} we briefly discuss alternative ways to determine \emph{all} (not necessarily obvious) fibration structures in this context. Finally in Section \ref{conc} we provide our conclusions and some possible future directions for research.

\section{CICYs, Fibrations, Symmetries, and Quotients}\label{review_sec}
In the following subsections we will provide a rapid review of the necessary geometric ingredients for the present study. Our focus will be on properties of Calabi-Yau three-folds defined as complete intersections in products of project spaces. On these covering spaces we will consider genus one fibrations, as well as discrete automorphisms, $\Gamma$, of the Calabi-Yau manifold, which we will use to construct non-simply connected Calabi-Yau three-folds defined as quotients, $q: X \to X/\Gamma$.

\subsection{CICY Three-Folds}
In this paper we will consider Calabi-Yau three-folds constructed as complete intersections in products of projective spaces (CICYs). The notation employed in our discussion will largely follow that of the original papers on the subject~\cite{Hubsch:1986ny,Green:1986ck,Candelas:1987kf,Candelas:1987du} and the text~\cite{Hubsch:1992nu}.

A family of CICYs is defined by a configuration matrix of the following form.
\begin{eqnarray} \label{conf1}
 [{\bf n}|{\bf q}] \equiv \left[\begin{array}{c|ccc}n_1 & q^1_1&\dots&q^1_K\\ \vdots & \vdots&\ddots&\vdots\\ n_m & q^m_1&\hdots&q^m_K\\\end{array}\right] 
\end{eqnarray}
In this matrix the entries $q_{\alpha}^r$ are positive semi-definite and the $n_{\alpha}$ are strictly positive (see \cite{Anderson:2015iia} for information on how these constraints can be relaxed). The first column of (\ref{conf1}) specifies the dimension of the projective spaces, the product of which forms the ambient manifold. The remaining columns then specify the degree of the $K$ defining relations of the complete intersection, with respect to the homogeneous coordinates of the projective space factors. In what follows we will label these defining relations $p_{\alpha}$. Note that in the above we have used the indices $r,s,\ldots=1,\ldots,m$ to label the ambient projective space factors $\mathbb{P}^{n_r}$ and the indices $\alpha,\beta \ldots = 1,\ldots ,K$ to label the defining relations $p_{\alpha}$.

For the matrix (\ref{conf1}) to define a three-fold we require that $3=\sum_{r=1}^m n_r -K$, given that the manifold is a complete intersection. The Calabi-Yau condition of vanishing first Chern class is satisfied if,
\begin{eqnarray}
 \sum_{\alpha = 1}^K q_\alpha^r = n_r + 1 \;.
\end{eqnarray}

It is clear that a configuration matrix of the form (\ref{conf1}) defines a family of manifolds in that it only specifies the degree of the defining relations and not specific polynomials. The possible choices of coefficients in the defining polynomials form a redundant description of, in general part of, the complex structure moduli space of the CICY. It can be shown that, for sufficiently general choices of complex structure, any configuration matrix defines a smooth complete intersection \cite{Green:1986ck,Hubsch:1992nu}.

\vspace{0.1cm}

It is clear that there are an infinite number of configuration matrices (\ref{conf1}) that describe CICY three-folds. In particular, one can keep making the ambient space bigger, while increasing the number of constraints $K$, in order to maintain a three dimensional space. Nevertheless, the set of Calabi-Yau three-folds described by the CICY construction is finite. Many different configuration matrices can describe the same Calabi-Yau manifold, and if these redundancies in description can be removed to a sufficient degree then an exhaustive classification can be made of the dataset. In particular there exists a set of 7,890 configuration matrices such that any three-fold of this type is described by at least one such matrix \cite{Candelas:1987kf}. An alternative complete list of three-fold configuration matrices has recently been derived wherein more of the properties of the Calabi-Yau manifolds in question descend in a simple way from the ambient space, leading to a higher degree of computational control \cite{Anderson:2017aux}. The CICY four-folds have also been classified in a similar manner \cite{Gray:2013mja,Gray:2014fla}. The topology (Hodge numbers, Chern classes, etc.) of CICY three-folds is well known (see for example \cite{Hubsch:1992nu}) and can be readily computed from the integer data in the configuration matrices given in \eref{conf1}.

\subsection{Discrete Symmetries and Quotients of CICY Three-Folds}\label{quots}

All of the CICY n-folds can be shown to be simply connected \cite{Hubsch:1992nu}. For odd dimensional cases, non-simply connected Calabi-Yau manifolds can be obtained from this construction by quotienting by an appropriate freely acting symmetry. Those freely acting symmetries which descend from a linear action on the homogeneous coordinates of the ambient spaces in the original CICY list of 7,890 matrices, and for which the symmetry restricted defining relation still generically leads to a smooth Calabi-Yau three-fold, have been classified in \cite{Braun:2010vc} (see \cite{Candelas:2008wb,Candelas:2010ve,Candelas:2015amz,Candelas:2016fdy,Constantin:2016xlj} for studies of the properties of these manifolds). 

We will begin with an example. Consider the following Calabi-Yau three-fold (\#6826 in the standard list \cite{cicylist}).
\begin{eqnarray} \label{notbicubic}
X = \left[ \begin{array}{c|ccc} \mathbb{P}^1 & 0&0&2 \\ \mathbb{P}^1 &0&0&2 \\ \mathbb{P}^4 & 2&2&1\end{array} \right]
\end{eqnarray}
This CICY's ambient space admits the following linear action on its homogeneous coordinates.
\begin{eqnarray} \label{symact1}
g:x_{i,a} &\to& (-1)^{a+1} x_{i,a}\\ \nonumber
g:y_{A} &\to& (-1)^{A+1} y_A
\end{eqnarray}
Here the $x_{i,a}$'s are the homogeneous coordinates on the two $\mathbb{P}^1$ factors with $i=1,2$ labeling the projective factors and $a=0,1$ labels the homogeneous coordinates on each of those two $\mathbb{P}^1$'s. The $y_A$ are the homogeneous coordinates on the $\mathbb{P}^4$ factor of the ambient space. This symmetry action has fixed points on $\mathbb{P}^1 \times \mathbb{P}^1 \times \mathbb{P}^4$ ambient space but these generically miss the Calabi-Yau hypersurface. The generic set of defining relations that are invariant under (\ref{symact1}), if one takes the final equation to transform with an over all negative sign, leads to a  smooth variety. Thus defining such a quotient of the complete intersection described by (\ref{notbicubic}) by the symmetry action (\ref{symact1}) leads to a smooth Calabi-Yau three-fold $\hat{X}$ with $\pi_1(\hat{X})=\mathbb{Z}_2$.

Given a quotient and its covering space $q: X \to \hat{X}=X/\Gamma$, the Chern classes and triple intersection numbers of $\hat{X}$ can be readily computed in terms of those associated to $X$. In general, a bundle $V$ on $X$ is called \emph{equivariant} if it satisfies $V= q^*(\hat{V})$ for some bundle $\hat{V}$ on $\hat{X}$. The bundle-valued cohomology of $\hat{V}$ on $X/\Gamma$ is precisely the $\Gamma$-invariant part\footnote{Here invariance is relative to the group action induced from the \emph{equivariant structure} on the bundle $V$. We will not go into details here and refer the reader to \cite{Donagi:2004ub,Anderson:2009mh} for a full disussion.} of that on $X$. That is,
\beq
\label{coho_descent} 
H^i(\hat{X}, \hat{V})=H^{i}_{inv} (X, V) \;.\eeq
For example, applying this to the holomorphic tangent bundle $V=TX$ yields the Hodge numbers of $\hat{X}$. The Hodge numbers of the CICY quotients of the form described above were calculated in \cite{Candelas:2008wb,Candelas:2010ve,Candelas:2015amz,Candelas:2016fdy,Constantin:2016xlj}. 

Chern classes have the following simple property under pull-back maps
\beq
c_i(q^*(\hat{V}))=q^*(c_i(\hat{V})) \;.
\eeq 
Labeling an integral basis of $H^2(X, \mathbb{Z})$ by $J_r$, where
$r=1\ldots h^{1,1}(X)$, and letting $\hat{J}_{a}$ be the generators of
$H^2(\hat{X}, \mathbb{Z})$, with $a=1,\ldots h^{1,1}(\hat{X})$, the relationship (\ref{coho_descent}) above implies that we can express the relationship between these sets of basis forms as
\beq\label{Js} 
q^*(\hat{J}_a)=K^{r}_{a} J_r \;, 
\eeq for some matrix
of integers $K^{r}_{a}$. Thus for an equivariant bundle $V=q^*(\hat{V})$, 
\beq
c_1(q^*(\hat{V}))^r J_r=c_1(q^*(\hat{V}))\\
=q^*(c_1(\hat{V})^a \hat{J}_a)=c_1(\hat{V})^a K^{r}_{a} J_r \;.  
\eeq
The coefficients of the first Chern classes of $V$ and
$\hat{V}$ are related then as
\beq 
c_1(q^*(\hat{V}))^r=c_1(\hat{V})^a
K^{r}_{a}~. 
\eeq
Likewise, a similar analysis \cite{Anderson:2009mh} yields that
\beq
c_{2}(\hat{V})^{cb}\hat{d}_{cba}=\frac{1}{|\Gamma|} c_2(q^*(\hat{V}))^{ts}d_{tsr}K^{r}_a
\eeq
where $d_{tsr}$ and $\hat{d}_{cba}$ are the triple intersection numbers on $X$ and $\hat{X}$ respectively. Similarly, the third Chern class has the following simple relationship
\beq
\int_X
c_3(q^*(\hat{V}))=\int_X q^*(c_3(\hat{V}))=|\Gamma| \int_{X/\Gamma} c_3(\hat{V})
\eeq
Finally, the triple intersection numbers of $X$ are defined as $d_{rst}=\int_{X} J_{r} \wedge J_{s} \wedge J_{t}$. 
\beq\label{triple}
 \int_X q^*(\hat{J}_a)\wedge q^*(\hat{J}_b) \wedge q^*(\hat{J}_c) =\int_X
q^*( \hat{J}_a\wedge \hat{J}_b\wedge \hat{J}_c) =|\Gamma| \int_{X/\Gamma} (\hat{J}_a\wedge \hat{J}_b\wedge \hat{J}_c) \;.\eeq
Expanding both sides and using \eref{Js} leads to
\beq\label{intersec_down}
 K_{a}^{r}K_{b}^{s}K_{c}^{t}d_{rst}=|\Gamma| \int_{X/\Gamma} (\hat{J}_a\wedge \hat{J}_b\wedge \hat{J}_c)=|\Gamma|\hat{d}_{abc}
 \eeq
We will illustrate these formulae for explicit examples in the following sections.

\subsection{Fibrations of CICY Three-Folds}

Extensive recent studies have shown that nearly all of the CICYs are multiply genus one fibered \cite{Gray:2014fla,Anderson:2015yzz,Anderson:2016ler,Anderson:2016cdu,Anderson:2017aux}. Two types of fibration have been studied in this context. The first are ``obvious" fibrations, that is, fibrations that can be observed to be present purely by a simple examination of the configuration matrix. The second type of fibrations that have been studied, which form an exhaustive set, are those classified by divisors satisfying certain properties: so called Koll\'ar fibrations.

\vspace{0.1cm}

Obvious fibrations of a CICY can be observed by simple manipulations of configuration matrices such as (\ref{notbicubic}). It is possible to perform arbitrary row and column permutations on a configuration matrix without changing the geometry that is described. Row permutations correspond to a simple reordering of the $\mathbb{P}^{n_r}$ ambient space factors while column permutations are a relabeling of the defining equations. Using such operations, one can ask whether the configuration matrix can be put in the following form:
\begin{equation}\label{mrfib}
X_{\textnormal{obv}}= \left[\begin{array}{c|c:c}  {\cal A}_1 & 0 & {\cal F} \\ \hdashline
{\cal A}_2 & {\cal B} & {\cal T} \end{array}\right] .
\end{equation}
In (\ref{mrfib}), ${\cal A}_1$ and ${\cal A}_2$ are  products of projective spaces and ${\cal F}, {\cal B}$ and ${\cal T}$ are block sub-matrices. If a configuration matrix can be put in such a form then the associated manifold can be described as a fibration of the variety described by $\left[ {\cal A}_1 | {\cal F}\right]$ over a base space described by $\left[ {\cal A}_2 | {\cal B}\right]$ with the ``twisting" of the fibre over the base is determined by the matrix ${\cal T}$ \cite{Gray:2014fla}. The fiber in such a case is always Calabi-Yau, and thus in the case that the fiber is one dimensional this corresponds to a torus fibration of the original manifold. In fact, a given configuration matrix can often be put in the form (\ref{mrfib}) in multiple inequivalent ways. Such analysis shows that the vast majority of CICY n-folds can be written in multiple inequivalent ways as Calabi-Yau fibrations of different types \cite{Gray:2014fla,Anderson:2017aux}. Note that it is important to check that the Calabi-Yau fiber is connected as it is possible to obtain, for example, fibers that are multiple tori embedded differently within the ambient space \cite{Anderson:2017aux}. Only connected fibers are considered in this work.

As an example of the above, the configuration matrix given in (\ref{notbicubic}) exhibits a rather sparse torus fibration structure and can only be written as an obvious genus one fibration in two inequivalent ways
\begin{eqnarray} \label{notbicubicfibs}
X= \left[ \begin{array}{c|ccc} \mathbb{P}^4 & 2 & 2&1 \\ \hdashline \mathbb{P}^1 &0&0&2 \\ \mathbb{P}^1 &0&0&2 \end{array}\right]\;~\rm{and}~\;X=\left[ \begin{array}{c|cc:c} \mathbb{P}^1 & 0 & 0&2 \\  \mathbb{P}^1 &0&0&2 \\ \hdashline\mathbb{P}^4 &2&2&1 \end{array}\right]~.
\end{eqnarray}
Note that in the first case above the matrix block ${\cal B}$ in (\ref{mrfib}) is trivial and thus the base of the fibration is simply $\mathbb{P}^1 \times \mathbb{P}^1$. In the second case the base is described by the configuration matrix $\left[ \begin{array}{c|cc} \mathbb{P}^4&2& 2\end{array} \right]$, which is a representation of $dP_5$.

\vspace{0.1cm}

All of the fibrations of a CICY, not just the obvious ones, can be obtained if one has enough control over the divisor and cone structure of the manifold \cite{kollar-criteria,ogu,wil,Anderson:2017aux}. For a Calabi-Yau n-fold, the existence of a genus-one fibrations has been conjectured by Koll\'ar to be determined by the following criteria \cite{kollar-criteria}.

\vspace{0.2cm}
\noindent{\it Conjecture: Let X be a Calabi-Yau n-fold. Then $X$ is genus-one fibered if and only if there exists a $(1,1)$-class $D$ in $H^2(X,\mathbb{Q})$ such that} 
\begin{align}\label{kollar}
&D\cdot C \geq 0~ \text{for every algebraic curve}~ C \subset X \nn \\
&D^{\textnormal{dim}(X)}=0  \\
&D^{\textnormal{dim}(X)-1} \neq 0 \nn \;.
\end{align}

\vspace{0.2cm}

For Calabi-Yau three-folds, it can be shown that one only need consider effective divisors in the above and in this case Oguiso and Wilson have proven that these conditions are sufficient to find all fibrations \cite{ogu,wil}. The conditions in Koll\'ar's conjecture have a clear intuitive origin. We can think of $D$ as a divisor in the base of the fibration such that $D^{\textnormal{dim}(X)-1}$ gives a point in that base. This then picks out the fiber of the fibration for us when these divisors are pulled back to the total space.

In general the obvious fibrations are a subset of these Koll\'ar fibrations. For example, the configuration (\ref{notbicubic}), while only exhibiting the $2$ obvious fibrations given in (\ref{notbicubicfibs}), in fact has $414$ fibrations in total \cite{Anderson:2017aux}.

\subsection{Fibrations in CICY Quotients}\label{fibs_quot}

In this section we will study the compatibility of fibrations of CICYs and quotients of them by freely acting discrete groups. In particular, one might be interested in which obvious fibrations exist in the quotiented, non simply-connected, geometry. Consider a symmetry action on a manifold described by a configuration matrix of the form (\ref{mrfib}). If the action is ``block diagonal", in that it does not transform coordinates in ${\cal A}_1$ into those of ${\cal A}_2$ and vice versa, then fiber and base directions are not being mixed and one might think that the fibration survives a quotienting of the manifold by the symmetry group. To see that this is true is straightforward. Firstly, the projection map before quotienting simply consists of deleting the coordinates of ${\cal A}_1$ in describing a point on the manifold that solves the defining relations. This projection map is compatible with a symmetry action of the form described above in that points on the total space of the manifold that are related to each other by the symmetry action project to points on the base which are also related by the restriction of the symmetry action to that space. Thus in grouping points on the manifold into equivalence classes related by the symmetry when taking the quotient, the original projection induces a natural such map on the quotient. Thus the following diagram is commutative and the quotiented manifold is a well defined fibration.
\begin{align}
\begin{array}{lll}
X& \stackrel{g}{\longrightarrow}& \hat{X} \\
\pi \downarrow&&\downarrow \hat{\pi} \\
B &\stackrel{\tilde{g}}{\longrightarrow}&\hat{B} \, .
\end{array}
\end{align}
In the above $\pi$ denotes the projection map on the upstairs manifold, $\hat{\pi}$ denotes the induced projection map on the quotiented manifold, $g$ denotes the quotient by the group action on the upstairs manifold and $\tilde{g}$ denotes the quotient of the induced group action on the base.

The question remains as to what the generic fiber looks like for $\hat{X}$. Fortunately this is easy to decide. Since we are quotienting by a finite group, the dimension of the fiber is obviously unchanged. Since the total space, for an odd dimensional Calabi-Yau manifold, is still a Calabi-Yau n-fold, the fibers must be Calabi-Yau as well. Thus, if the upstairs space was fibered by Calabi-Yau m-folds, so too is the quotiented manifold. In particular, if we start with a torus fibration then the quotient will be torus fibered too. Note that an argument of this type works quite generally, including for non-complete intersections of the form (\ref{mrfib}) in arbitrary ambient spaces ${\cal A}_1$ and ${\cal A}_2$ with an appropriate block diagonal group action.

For the obvious fibrations of the example given in (\ref{notbicubic}) the fibration structure is clearly preserved under quotienting by the action (\ref{symact1}). The group action acts within projective space ambient factors and does not transform one $\mathbb{P}^{n_r}$ into another. Thus this symmetry action is indeed of the block diagonal form described above. It should be noted that the base of the fibrations obtained in this manner are singular in general, being $\mathbb{P}^1 \times \mathbb{P}^1/\mathbb{Z}_2$ and $dP_5/\mathbb{Z}_2$ respectively in this case. The total space is still smooth by construction, however. This is possible due to multiple fibers appearing over the orbifold singularities in the base \cite{Anderson:2018heq}.

As one further illustration, consider the tetra-quadric three-fold defined as a single hypersurface with Hodge numbers $(h^{1,1},h^{2,1})=(2,68)$
\begin{eqnarray} \label{tetraquad}
X = \left[ \begin{array}{c|c} \mathbb{P}^1 &2 \\ 
\mathbb{P}^1 &2 \\
\mathbb{P}^1 & 2 \\
\mathbb{P}^1 & 2
\end{array} \right]~.
\end{eqnarray}
This geometry exhibits several different descriptions as a genus one fibration, $\pi: X \to \mathbb{P}^1 \times \mathbb{P}^1$. For illustration, let us take the first two rows to describe a genus one fiber of multi-degree $\{2,2\}$ in $\mathbb{P}^1 \times \mathbb{P}^1$ and the latter two columns to give rise to the base $\mathbb{P}^1 \times \mathbb{P}^1$.

With that fiber/base in mind, we can consider discrete symmetry actions on $X$ and their relationship with the fibration chosen above. For example, the generators of two different symmetries, acting on the eight projective coordinates of the four ambient $\mathbb{P}^1$ factors are given by
\beq
\mathbb{Z}_2: \left(
\begin{array}{cccccccc}
 -1 & 0 & 0 & 0 & 0 & 0 & 0 & 0 \\
 0 & 1 & 0 & 0 & 0 & 0 & 0 & 0 \\
 0 & 0 & -1 & 0 & 0 & 0 & 0 & 0 \\
 0 & 0 & 0 & 1 & 0 & 0 & 0 & 0 \\
 0 & 0 & 0 & 0 & -1 & 0 & 0 & 0 \\
 0 & 0 & 0 & 0 & 0 & 1 & 0 & 0 \\
 0 & 0 & 0 & 0 & 0 & 0 & -1 & 0 \\
 0 & 0 & 0 & 0 & 0 & 0 & 0 & 1 \\
\end{array}
\right)~~~\textnormal{and}~~\mathbb{Z}_4: \left(
\begin{array}{cccccccc}
 0 & 0 & 0 & 0 & 1 & 0 & 0 & 0 \\
 0 & 0 & 0 & 0 & 0 & -1 & 0 & 0 \\
 0 & 0 & 0 & 0 & 0 & 0 & 1 & 0 \\
 0 & 0 & 0 & 0 & 0 & 0 & 0 & 1 \\
 1 & 0 & 0 & 0 & 0 & 0 & 0 & 0 \\
 0 & -1 & 0 & 0 & 0 & 0 & 0 & 0 \\
 0 & 0 & 1 & 0 & 0 & 0 & 0 & 0 \\
 0 & 0 & 0 & 1 & 0 & 0 & 0 & 0 \\
\end{array}
\right) \;.
\eeq
In the first case of the given $\mathbb{Z}_2$ symmetry, the decomposition between fiber and base described above is preserved and as argued above, this fibration will descend to the quotient geometry. However, for the $\mathbb{Z}_4$ symmetry on the other hand, the chosen fiber and base are non-trivially identified under the discrete group action, since the symmetry relates points in the first and third (and second and forth) ambient $\mathbb{P}^1$ factors. In this latter case, this particular fibration structure is not preserved in passing to the quotient manifold, $\hat{X}=X/\mathbb{Z}_4$.

In the rest of this paper we will classify which obvious fibrations of the CICY three-folds are preserved by which of the freely acting symmetries on those manifolds descending from linear actions on the ambient space. It is important to note in this context that we will be using the original symmetry classification of \cite{Braun:2010vc}. As such new possible linear actions associated to the new ``maximally favorable CICY list" found in \cite{Anderson:2017aux} are not considered. We will also discuss the preservation of more general Koll\'ar fibrations in this context.

\section{Results: Obvious fibrations} \label{res1}

Of the 7,890 CICY configuration matrices in the standard list \cite{Candelas:1987kf}, $195$ have symmetries that descend from linear actions on the ambient space \cite{Braun:2010vc}. These $195$ configuration matrices exhibit a total of 1,695 symmetries and 1,600 fibrations. The number of symmetry-fibration pairs is 20,700. This is the data set with which we shall work.

\begin{figure}[!h]\centering
\includegraphics[width=0.69\textwidth]{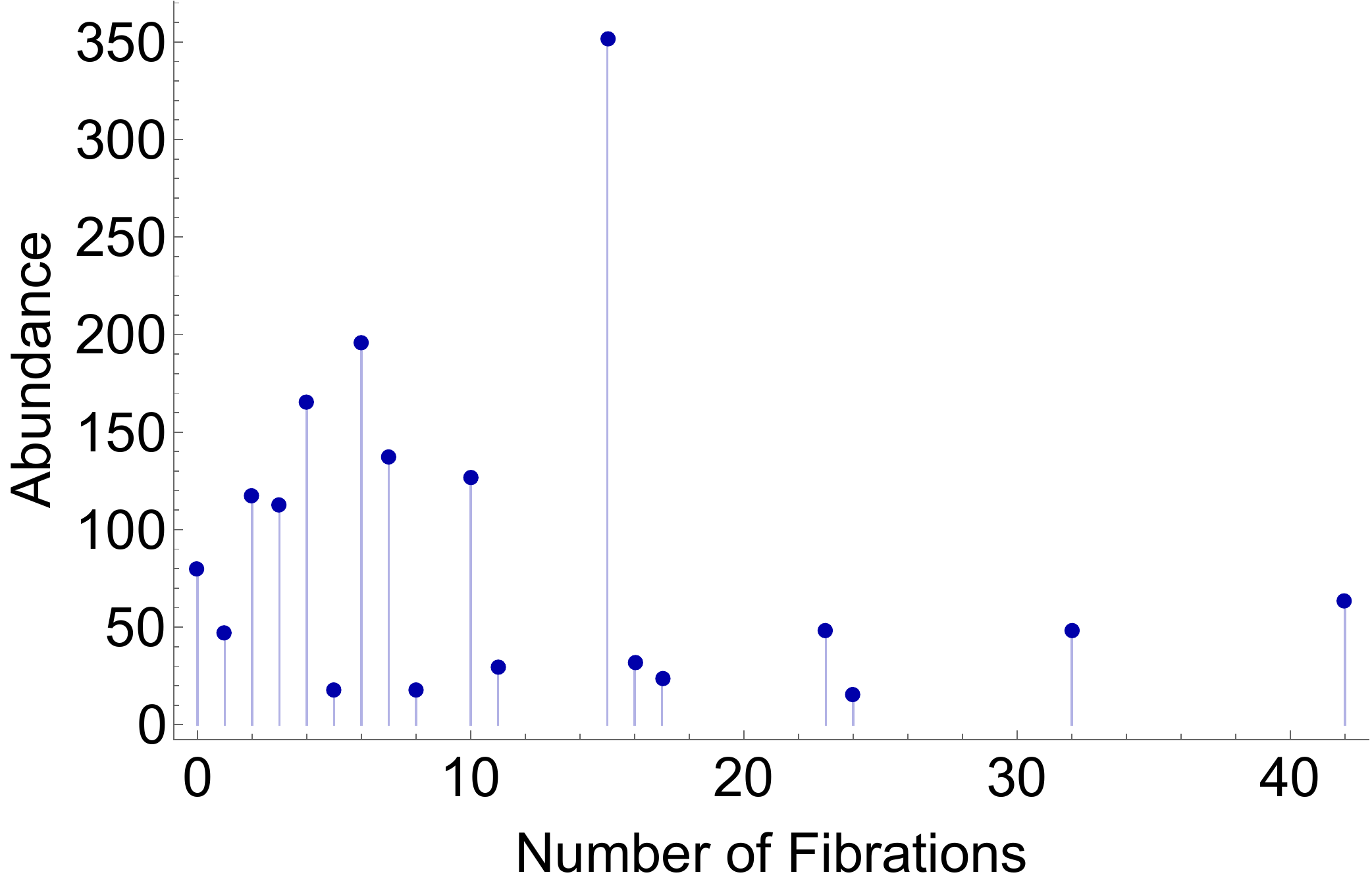}
\caption{{\it The frequency (``Abundance") with which a non-simply connected Calabi-Yau manifold with a given number of obvious genus one fibrations appears in the data set described in the text.}}
\label{fig3}
\end{figure}

Of the 20,700 fibration-symmetry pairs, 17,161, that is $83\%$, are compatible in the sense that quotienting by the symmetry preserves the fibration. Of the 1,600 fibrations on the original upstairs CICYs, all but $361$, that is $77\%$ are preserved on quotienting by at least one symmetry. Of the 1,695 quotient CICYs, $63$ are not fibered upstairs. Of the remaining 1,632, all but $80$, that is $95\%$ preserve at least one obvious fibration. A plot of the number of quotiented CICYs exhibiting a given number of fibrations can be found in Figure \ref{fig3}. It is interesting to note that all quotients, $\hat{X}$, which have the same value for $h^{1,1}$ as the upstairs parent manifold, $X$, as computed using the techniques in \cite{Constantin:2016xlj}, preserve all obvious fibrations. This might be expected given the discussion of when fibrations are preserved in the previous section and the fact that symmetries that do not reduce $h^{1,1}$ tend to have an action that is block diagonal in the projective space ambient factors.
 
The number of fibrations preserved depends strongly on the type of symmetry being quotiented by. In Table \ref{fig1} we give the number of fibration/symmetry pairs for each type of symmetry present and how many of these pairs are compatible. It is interesting to note that no non-abelian symmetries preserve any fibrations in the list. In addition, any single factor above rank 6 results in no fibrations being preserved. This result is compatible with known hints towards the maximal degree of quotient singularities in the base of F-theory models \cite{Bhardwaj:2015oru,deBoer:2001wca}.
\begin{table}[t]
 \begin{center}$
\begin{array}{|c|c|c|} \hline \textnormal{Symmetry} & \textnormal{Abundance} & \textnormal{Fibration Compatible} \\ \hline\hline \mathbb{Z}_2 & 9276 & 8812 \\\hline
\mathbb{Z}_3 & 376 & 175 \\\hline
\mathbb{Z}_4 & 364 & 120 \\\hline
\mathbb{Z}_5 & 30 & 0 \\\hline
\mathbb{Z}_6 & 500 & 62 \\\hline
\mathbb{Z}_8 & 32 & 0 \\\hline
\mathbb{Z}_{10} & 20 & 0 \\\hline
\mathbb{Z}_{12} & 52 & 0 \\\hline
\mathbb{Z}_2 \times \mathbb{Z}_2 & 9199 & 7711 \\\hline
\mathbb{Z}_3 \times \mathbb{Z}_3 & 176 & 176 \\\hline
\mathbb{Z}_4 \times \mathbb{Z}_2 & 305 & 105 \\\hline
\mathbb{Z}_4 \times \mathbb{Z}_4 & 30 & 0 \\\hline
\mathbb{Z}_8 \times \mathbb{Z}_2 & 48 & 0 \\\hline
\mathbb{Q}_8 & 90 & 0 \\\hline
\mathbb{Q}_8 \times \mathbb{Z}_2 & 62 & 0 \\\hline
\mathbb{Z}_3 \rtimes \mathbb{Z}_4 & 52 & 0 \\\hline
\mathbb{Z}_4 \rtimes \mathbb{Z}_4 & 48 & 0 \\\hline
\mathbb{Z}_8 \rtimes \mathbb{Z}_2 & 30 & 0 \\\hline
\mathbb{Z}_{10} \times \mathbb{Z}_2 & 10 & 0 \\\hline
\end{array}$
\end{center}
\caption{\emph{The number of fibrations that are compatible with each symmetry that appears in the Braun's classification \cite{Braun:2010vc}. ``Abundance" denotes the number of fibrations appearing in CICYs admitting a given symmetry, and ``Fibration Compatible" details how many of these fibrations are compatible with the symmetry and thus descend to the quotiented manifold \label{fig1}}}
\end{table}

In Table \ref{fig5} we present the base manifolds that appear in obvious fibrations of the quotient CICY three-folds, together with how many cases each individual base appears in in the list. Note that all of the bases are quotients of either $\mathbb{P}^1 \times \mathbb{P}^1$ or del Pezzo surfaces.

\begin{table}[t] \label{fig5}
\begin{center}
$\begin{array}{|c|c|} \hline \textnormal{Base} & \textnormal{Abundance} \\ \hline \hline 
(\mathbb{P}^1\times\mathbb{P}^1)/\mathbb{Z}_2  & 7589 \\ \hline
(\mathbb{P}^1\times\mathbb{P}^1)/\mathbb{Z}_2 \times \mathbb{Z}_2 & 6610 \\\hline
(\mathbb{P}^1\times\mathbb{P}^1)/\mathbb{Z}_4 & 72 \\\hline
(\mathbb{P}^1\times\mathbb{P}^1)/\mathbb{Z}_4 \times \mathbb{Z}_2 & 83 \\\hline
\mathbb{P}^2/\mathbb{Z}_3 & 90 \\\hline
\mathbb{P}^2/\mathbb{Z}_3 \times \mathbb{Z}_3 & 104 \\\hline
dP_3/\mathbb{Z}_2 & 912 \\\hline
dP_3/\mathbb{Z}_3 & 51 \\\hline
dP_3/\mathbb{Z}_6 & 62 \\\hline
dP_5/\mathbb{Z}_2 & 258 \\\hline
dP_5/\mathbb{Z}_4 & 18 \\\hline
dP_5/\mathbb{Z}_2\times \mathbb{Z}_2 & 1059 \\\hline
dP_6/\mathbb{Z}_3 & 16 \\\hline
dP_7/\mathbb{Z}_2 & 21 \\\hline
dP_9/\mathbb{Z}_2 & 32 \\\hline
dP_9/\mathbb{Z}_3 & 18 \\\hline
dP_9/\mathbb{Z}_4 & 30 \\\hline
dP_9/\mathbb{Z}_2\times \mathbb{Z}_2 & 42 \\\hline
dP_9/\mathbb{Z}_3 \times \mathbb{Z}_3 & 72 \\\hline
dP_9/\mathbb{Z}_4 \times \mathbb{Z}_2 & 22 \\\hline
\end{array}$
\end{center}
\caption{\emph{All of the base manifolds, $\hat{B}$ of genus one fibrations, $\hat{\pi}: \hat{X} \to \hat{B}$, which appear in the quotient CICY dataset being considered. Abundance refers to the number of different fibrations in which the associated base appears.} \label{fig5}}
\end{table}

In every case where a fibration is preserved, the group which appears in the quotient defining the base is precisely the same as the group appearing in the definition of the total space (and not a proper subgroup).

\vspace{0.2cm}

\begin{figure}[!h]\centering
\includegraphics[width=0.89\textwidth]{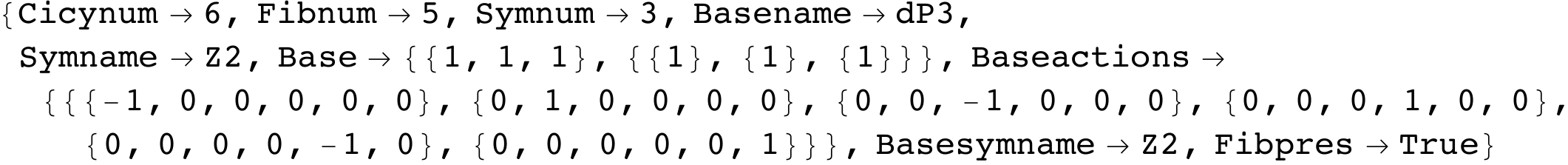}
\caption{{\it An example of the data describing a fibration of a quotiented CICY.}}
\label{thisoneisfirst}
\end{figure}

The data of the fibrations of the quotiented CICYs can be found at this web address \cite{quotientlist}. The data is presented in a Mathematica readable format in a plain text file. The data is in the form of a list of lists. Each entry corresponds to a given fibration/symmetry pair and is of the form shown in Figure \ref{thisoneisfirst}.
The first three entries give the CICY number of the parent configuration \cite{cicylist}, the fibration number according to the standard list \cite{thefiblist}, and the symmetry number given in the standard list \cite{cicylist}. The fourth entry gives the name of the surface which is quotiented to obtain the base of the fibration. The fifth entry gives the name of the symmetry being quotiented by, the sixth entry gives the description of the base that appears, first as a list of the dimensions of the projective space factors and then as a matrix giving the degrees of the defining relations. The ``Baseactions" entry describes how the symmetry acts on the base configuration and the ``Basesymname" entry describes the symmetry that the base is quotiented by. Finally, the last entry in the list says whether the fibration and symmetry are compatible, in other words whether this fibration of the CICY is associated with a fibration of the quotiented Calabi-Yau as well.

\begin{figure}[!h]\centering 
\includegraphics[width=0.89\textwidth]{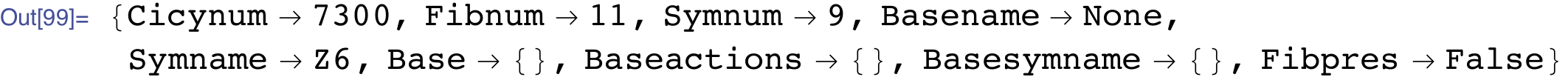}
\caption{{\it An example of the data describing a fibration of a quotiented CICY where the fibration and symmetry being considered are not compatible.}}\label{thisonenext}
\end{figure}

If the fibration is not compatible with the symmetry some of these entries will be the empty list, as shown in Figure \ref{thisonenext}, for obvious reasons.

Finally, we should mention that if the base is a simple description of $\mathbb{P}^1 \times \mathbb{P}^1$ or $\mathbb{P}^2$ and does not have any defining relations then the ``Base" entry will contain an empty matrix. Thus, in the case of a simple $\mathbb{P}^1 \times \mathbb{P}^1$ base we will have the entry $\textnormal{Base}\to \{\{1,1\},\{\{\}\}\}$ .

\section{Obvious vs. Koll\'ar fibrations: An Illustration} \label{res2}
In the previous section we have characterized some of the genus one fibrations of the dataset of 1,695 CY three-folds constructed as quotients of CICYs by freely acting discrete automorphisms. These so-called ``obvious" fibrations are clearly plentiful and natural to consider because of the covering space structure of $\hat{X}$. However, past experience \cite{Anderson:2017aux} has shown that for some manifolds, the obvious fibrations may in fact be a very small subset of the total number of such fibrations that exist. Indeed, for the CICY three-folds there are examples where the non-obvious fibrations number in the thousands, or in one case, even lead to a countably infinite set of fibrations. With that in mind, in this section we would like to turn our attention to the counting of obvious vs. Koll\'ar fibrations in CICY quotients, $q: X \to \hat{X}=X/\Gamma$. It is beyond the scope of this work to perform a full fibration scan along the lines of \cite{Anderson:2017aux} (as the structure of the K\"ahler and Mori cones of CICY quotients remains unknown in some cases), however we provide a simple case study below to illustrate the essential ideas.

Consider the following CICY three-fold
\begin{equation}\label{cicy6947}
X= \left[ \begin{array}{c|cccccccc} \mathbb{P}^1 & 1 & 1 &0 &0 & 0 &0  &0 & 0\\ 
\mathbb{P}^1 & 0 & 0 &1 &1 &0 &0 &0 &0\\ 
\mathbb{P}^1 & 0 & 0 &0 &0 &1 &1  &0 &0 \\
\mathbb{P}^1 & 0 & 0 &0 &0 &0 &0  &1 &1 \\
\mathbb{P}^7 &1 & 1 &1 &1 &1 &1  &1 &1 
 \end{array}\right]\;,
\end{equation}
This manifold has $(h^{1,1}(X),h^{2,1}(X))=(5, 37)$ and admits a freely acting $\mathbb{Z}_8$ symmetry. Labeling the coordinates of the four $\mathbb{P}^1$ factors as $x_i,y_i,z_i,w_i$ with $i=0,1$ and that of $\mathbb{P}^7$ as $u_a$ with $a=0, \ldots 7$, the generator of $\mathbb{Z}_8$ acts on the coordinates of the ambient product of projective spaces as 
\begin{align}\label{symm_act}
&x_i \to (-1)^{i}z_i ~~,~~y_i \to w_i~~,~~z_i \to y_i~~,~~w_i \to x_i \\
&u_a \to (-1)^a u_a ~~\text{for a=0,1} \\
&u_a \to (-1)^a e^{\frac{2\pi i }{4}}~~\text{for a=2,3} \\
&u_a \to (-1)^a e^{\frac{2\pi i }{8}}~~\text{for a=4,5} \\
&u_a \to (-1)^a e^{\frac{6\pi i }{8}}~~\text{for a=6,7}  \;.
\end{align}
This action is coupled to a non-trivial action on the defining equations (equivalently a non-trivial equivariant action on the normal bundle, $N=\bigoplus_i L_i$ with $i=1,\ldots 8$ associated to the columns of \eref{cicy6947}) given by
\begin{equation}
\left(
\begin{array}{cccccccc}
 0 & 0 & 0 & 0 & 1 & 0 & 0 & 0 \\
 0 & 0 & 0 & 0 & 0 & 1 & 0 & 0 \\
 0 & 0 & 0 & 0 & 0 & 0 & 1 & 0 \\
 0 & 0 & 0 & 0 & 0 & 0 & 0 & 1 \\
 0 & 0 & 1 & 0 & 0 & 0 & 0 & 0 \\
 0 & 0 & 0 & 1 & 0 & 0 & 0 & 0 \\
 1 & 0 & 0 & 0 & 0 & 0 & 0 & 0 \\
 0 & -1 & 0 & 0 & 0 & 0 & 0 & 0 \\
\end{array}
\right) \;.
\end{equation}

By defining the smooth quotient $\hat{X}=X/\mathbb{Z}_8$ associated to the symmetry given above, we arrive at a new geometry \cite{Braun:2010vc} with Hodge numbers \cite{Constantin:2016xlj} $(h^{1,1}(\hat{X}),h^{2,1}(\hat{X}))=(2,6)$.

It is straightforward to verify using the observations of Section \ref{review_sec} that this manifold admits six obvious fibrations. Furthermore, from \cite{Anderson:2017aux} it is known that these are in fact \emph{all} of the genus one fibrations of this geometry. Each of these six distinct fibrations takes the form $\pi: X \to \mathbb{P}^1 \times \mathbb{P}^1$. The six are obtained by considering all possible pairs of ambient space $\mathbb{P}^1$ factors in \eref{cicy6947} to be the base. For example, taking the first two $\mathbb{P}^1$ factors as the base yields a genus one fiber whose form is given by the CICY configuration matrix
\beq
\left[ \begin{array}{c|cccccccc} 
\mathbb{P}^1 & 0 & 0 &0 &0 &1 &1  &0 &0 \\
\mathbb{P}^1 & 0 & 0 &0 &0 &0 &0  &1 &1 \\
\mathbb{P}^7 &1 & 1 &1 &1 &1 &1  &1 &1 
 \end{array}\right]=\left[ \begin{array}{c|cccc} 
\mathbb{P}^1  &1 &1  &0 &0 \\
\mathbb{P}^1 &0 &0  &1 &1 \\
\mathbb{P}^3 &1 &1  &1 &1 
 \end{array}\right]\;,
\eeq
 After the analysis of Section \ref{res1} we find that \emph{none of these fibrations are compatible with the $\mathbb{Z}_8$ symmetry given above}. By inspection, this is not surprising since it is clear that the action given in \eref{symm_act} non-trivially ``glues" together all four of the ambient $\mathbb{P}^1$ factors, spoiling the fiber/base decomposition of each of the six fibrations.
 
However, with these observations in hand, it is natural to ask whether $\hat{X}$ could still be a genus one fibered geometry. Is it possible that the quotient manifold gains fibration structures not inherited from its covering space? We can answer this question by considering a full fibration analysis using the Koll\'ar criteria in \eref{kollar}. To this end we must consider a basis of divisors $\hat{D}_a$ on $\hat{X}$ and their intersection numbers. Such information will make it possible to determine whether any divisors satisfying the Koll\'ar criteria in \eref{kollar} exist. 

The Picard group of $\hat{X}$ is spanned by $\hat{D}_1,\hat{D}_2$ which are related to divisors on $X$ as
\begin{align}\label{basis_down}
&q^*(\hat{D}_1)=D_1 + D_2 + D_3 + D_4 \\
& q^*(\hat{D}_2)=D_5
\end{align}
where $D_r$, with $r=1,\ldots 5$ correspond to the hyperplanes associated to the projective space appearing in the $r$-th row of \eref{cicy6947}. Using the dual cohomology description this gives us a basis of K\"ahler $(1,1)$-forms $\hat{J}_a$ on $\hat{X}$.

Following the discussion of Section \ref{quots} we can compute the intersection numbers on $\hat{X}$. On $X$ itself the standard formulas (see e.g. \cite{Hubsch:1992nu}) yield
\begin{eqnarray}
d_{rst}=\{\{\{0, 0, 0, 0, 0\}, \{0, 0, 2, 2, 4\}, \{0, 2, 0, 2, 4\}, \{0, 2, 2, 0, 
   4\}, \{0, 4, 4, 4, 8\}\},  \\ \nonumber
  \{\{0, 0, 2, 2, 4\},
    \{0, 0, 0, 0, 0\}, \{2, 0, 0, 
   2, 4\}, \{2, 0, 2, 0, 4\}, \{4, 0, 4, 4, 8\}\}, \\ \nonumber \{\{0, 2, 0, 2, 4\}, \{2, 0, 
   0, 2, 4\}, \{0, 0, 0, 0, 0\}, \{2, 2, 0, 0, 4\}, \{4, 4, 0, 4, 8\}\}, \\ \nonumber\{\{0, 
   2, 2, 0, 4\}, \{2, 0, 2, 0, 4\}, \{2, 2, 0, 0, 4\}, \{0, 0, 0, 0, 0\}, \{4,
    4, 4, 0, 8\}\}, \\ \nonumber \{\{0, 4, 4, 4, 8\}, \{4, 0, 4, 4, 8\}, \{4, 4, 0, 4, 
   8\}, \{4, 4, 4, 0, 8\}, \{8, 8, 8, 8, 16\}\}\}
\end{eqnarray}
Then, applying the formulae of Section \ref{quots}, \eref{intersec_down} it is straightforward to verify that the triple intersection numbers of $\hat{X}$ in terms of the basis given in \eref{basis_down} are
\beq\label{eg_intersec}
\hat{d}_{abc}=\{\{\{6, 6\}, \{6, 4\}\}, \{\{6, 4\}, \{4, 2\}\}\}
\eeq

Now, with these pieces in place, one can ask whether or not there exist any divisors satisfying $\hat{D}^2 \neq 0$ and $\hat{D}^3=0$ where
\beq
\hat{D}=a \hat{D}_1 + b \hat{D}_2
\eeq
and $a,b \in \mathbb{Z}$. If a solution of this type can be obtained it must further be checked whether $\hat{D}$ is nef on $\hat{X}$. However, in this case the answer is straightforward. There are \emph{no non-trivial solutions} to the Koll\"ar criteria that $\hat{D}^2 \neq 0$ and $\hat{D}^3=0$ on $\hat{X}$ for the intersection numbers in \eref{eg_intersec}. As a result, we can conclusively state that ${\hat X}$ is not genus one fibered. In this case, the obvious fibration result gives a complete answer for the fibrations of $\hat{X}$.

Likewise, for each manifold/symmetry pair in consideration in this work, a comprehensive analysis of fibrations on $X/\Gamma$ is possible in principle. One reason to consider such a scan is to determine what percentage of this dataset of non-simply connected CY three-folds is genus one fibered. As described in Section \ref{res1} above, by considering the obvious fibrations alone, it is clear that this set of CY three-folds follows the same patterns observed within all known datasets of CY three-folds -- namely that nearly all of these manifolds admit multiple descriptions as a genus one fibration. In the case of these CICY quotients with $\pi_1 (\hat{X}) \neq 0$, the fraction of fibered geometries is $\geq 95\%$ from the obvious fibrations alone.

Within all known datasets, the remaining small set of manifolds that are not fibered are all geometries with small values of $h^{1,1}$. Intriguingly, within the set of 7,890 CICY covering spaces, the highest Hodge number observed for a geometry that is not fibered is $h^{1,1}(X)=4$. Within the present scan however, we find one apparent quotient manifold which admits no obvious fibrations with $h^{1,1}(\hat{X})=6$. In this case however, there is some reason to believe that an analysis like the one above will give a different answer and bring to light non-obvious fibrations. The three-fold in question is a $\mathbb{Z}_4$ quotient of a CICY with Hodge numbers $(h^{1,1}(X),h^{2,1}(X))=(19,19)$ -- the so-called ``Schoen" \cite{schoen} or ``split bi-cubic" three-fold \cite{Candelas:1987kf}. This manifold is in fact known to have an infinite number of inequivalent, non-obvious genus one fibrations (see e.g. \cite{Anderson:2017aux}). So it is certainly possible that one of these non-obvious fibrations descends to the quotient $\hat{X}$. Furthermore, there is an even stronger hint that this manifold may be genus one fibered. Within the CICY dataset there exists another $\mathbb{Z}_4$ quotient of the Schoen manifold which yields a quotient manifold with the same Hodge numbers of $(h^{1,1}(X),h^{2,1}(X))=(6,6)$ which does admit an obvious fibration. Interestingly this is not a proof that the original Schoen quotient is genus one fibered since even the same discrete group, acting on the same manifold could lead to a different quotient three-fold via a distinct linear realization in a CICY description. However, it is a strong hint that once again as $h^{1,1}$ increases past a low bound (of order $\sim h^{1,1}=4$) all non-simply connected three-folds in this set appear to admit genus one fibrations.

\section{Conclusions} \label{conc}
In this work we have systematically investigated genus one fibrations within a dataset of non-simply connected Calabi-Yau three-folds. The manifolds we consider are constructed as quotients of simply connected manifolds (constructed algebraically as complete intersections) by freely acting discrete automorphisms. The only dataset for which discrete automorphisms have been systematically studied is the 7,890 manifolds constructed as complete intersections in products of projective spaces. For this set, all freely acting discrete symmetries realized via a linear action on the coordinates of the ambient product of projective spaces have been classified \cite{Braun:2010vc}. Using this set of manifolds we construct the smooth quotient geometries, $q: X \to \hat{X}=X/\Gamma$, and investigate the presence of genus one fibrations. We have produced a dataset of 17,161 ``obvious" genus one fibrations which descend from fibrations of the simply connected covering spaces.

There are a number of open questions leading on from this study and aspects of physical interest that it would be useful to explore further. We list a few of these below:

\begin{itemize}
\item It would be interesting to extend a study such as this to a larger set of non-simply connected Calabi-Yau three-folds. At present though, such examples remain rare. A recent study of linearly realized discrete symmetry actions on a subset of the toric hypersurfaces \cite{Braun:2017juz} yielded only a handful of manifolds beyond those considered here. It would be fruitful to potentially extend this set further to include manifolds constructed via non-Abelian GLSMs (see e.g. \cite{Hori:2006dk,Jockers:2012zr,Hori:2013gga,Caldararu:2017usq}) the recent gCICY construction \cite{Anderson:2015iia}.
\item Here we have focused primarily on genus one fibrations of CY three-folds. Our techniques could be similarly applied to study $K3$ (or Abelian surface) fibrations of CY manifolds. This would be interesting to study in the context of non-simply connected CY three-folds as such manifolds remain largely unexplored in the context of string dualities. In particular, heterotic/F-theory duality on these backgrounds (in compactifications to either 6- or 4-dimensions) could yield novel structure.
\item In this work we have not attempted to characterize which genus one fibered geometries also admit a section (i.e. which are also \emph{elliptically} fibered). Naively it seems that \emph{all non-simply connected CY manifolds constructed as smooth quotients in this way will be genus one fibered} (i.e. fail to have a section to the fibration). This expectation arises from the fact that generically the base manifolds within this set will be singular\footnote{Note that an exception to this general observation is the case when the base manifold is an Enriques surface (see e.g. \cite{gross_finite}.}. In order to preserve the smoothness of the total CY three-fold, multiple fibers have been observed over the singular points in the base. It is known that such isolated multiple fibers are incompatible with the existence of a section \cite{gross_mult}. Certainly within the dataset in hand, cursory inspections fail to find elliptically fibered quotient manifolds. However, it would be good to study these properties more carefully and prove the statements above in general.
\item It would be interesting to consider the quotient geometries described here as potential backgrounds for T-brane solutions \cite{Cecotti:2010bp,Donagi:2011jy,Anderson:2017zfm,Anderson:2017rpr,Anderson:2013rka}. In particular the role of $\pi_1(X) \neq 0$ in this context would be interesting to explore.
\item Perhaps most importantly, the manifolds/fibrations cataloged here have recently been shown to provide novel new vacua for F-theory compactifications to 6-dimensions \cite{Anderson:2018heq} . The pairing of smooth CY total spaces with singular base manifolds bring to light the relationship between geometric features (including multiple fibers and orbifold singularities in the base manifold) and effects in the physical theory (such as superconformal loci and discretely charged superconformal matter). 

\vspace{7pt}

Within this context, a number of open questions remain. It was observed in \cite{Anderson:2018heq} that the Tate-Schaferavich group (more generally the group of CY torsors) associated to the non-simply connected genus one fibered manifolds is frequently linked to the order of the discrete group $\Gamma$ used to construct the geometries as quotients. However the exact relationship remains unexplored. Furthermore, it is expected that the number of superconformal loci that can be coupled to a supergravity theory (realized as orbifold singularities in the base of a compact CY manifold) and the discrete charges of the associated matter fields are bounded. To date, the exact bounds are not yet known. We hope that this dataset may shed further light on these questions
\end{itemize}
 
 It is our hope that this dataset will provide an explicit playground in which to investigate these and other questions. We plan to continue exploring the physics associated to such geometries in future work.

\section*{Acknowledgments}

The work of L.A. and J.G. is supported in part by NSF grant PHY-1720321. This research is part of the working group activities of the the 4-VA initiative ``A Synthesis of Two Approaches to String Phenomenology". 




\begin{thebibliography}{99}
\ifx\doiref\asklfhas\newcommand{\doiref}[2]{\href{http://dx.doi.org/#1}{#2}}\fi
\raggedright 
\ifx\arxivref\asklfhas\newcommand{\arxivref}[2]{\href{http://arxiv.org/abs/#1}{arXiv:#1}}\fi
\raggedright

\bibitem{quotientlist}
The data produced during the course of preparing this paper can be found here: http://www1.phys.vt.edu/quotientdata/.

\bibitem{Vafa:1996xn} 
  C.~Vafa,
  ``Evidence for F theory,''
  Nucl.\ Phys.\ B {\bf 469}, 403 (1996)
  doi:10.1016/0550-3213(96)00172-1
  [hep-th/9602022].

\bibitem{gross_finite}
M.~Gross,
``A Finiteness Theorem for Elliptic Calabi-Yau Threefolds",
Duke Math. J. 74, no. 2, (1993), 271. 
arXiv:alg-geom/9305002.

  \bibitem{2016arXiv160802997D}
  G.~Di Cerbo and R. Svaldi,
  ``Log birational boundedness of Calabi-Yau pairs"
  arXiv:1608.02997.


\bibitem{grassi}
A.~Grassi,
``On minimal models of elliptic threefolds",
Matth. Ann. 290. (1991), 287.


\bibitem{Rohsiepe:2005qg} 
  F.~Rohsiepe,
  ``Fibration structures in toric Calabi-Yau fourfolds,''
  hep-th/0502138.
  
\bibitem{Johnson:2014xpa} 
  S.~B.~Johnson and W.~Taylor,
  ``Calabi-Yau threefolds with large $h^{2,1}$,''
  JHEP {\bf 1410}, 23 (2014)
  [arXiv:1406.0514 [hep-th]].
  
  
\bibitem{Johnson:2016qar} 
  S.~B.~Johnson and W.~Taylor,
  ``Enhanced gauge symmetry in 6D F-theory models and tuned elliptic Calabi-Yau threefolds,''
  Fortsch.\ Phys.\  {\bf 64}, 581 (2016)
  [arXiv:1605.08052 [hep-th]].
  
\bibitem{Candelas:2012uu} 
  P.~Candelas, A.~Constantin and H.~Skarke,
  ``An Abundance of K3 Fibrations from Polyhedra with Interchangeable Parts,''
  Commun.\ Math.\ Phys.\  {\bf 324}, 937 (2013)
  [arXiv:1207.4792 [hep-th]].
  
  
  \bibitem{Gray:2014fla} 
  J.~Gray, A.~S.~Haupt and A.~Lukas,
  ``Topological Invariants and Fibration Structure of Complete Intersection Calabi-Yau Four-Folds,''
  JHEP {\bf 1409}, 093 (2014)
  doi:10.1007/JHEP09(2014)093
  [arXiv:1405.2073 [hep-th]].

    \bibitem{Anderson:2016cdu} 
  L.~B.~Anderson, X.~Gao, J.~Gray and S.~J.~Lee,
  ``Multiple Fibrations in Calabi-Yau Geometry and String Dualities,''
  JHEP {\bf 1610}, 105 (2016)
  doi:10.1007/JHEP10(2016)105
  [arXiv:1608.07555 [hep-th]].
  
  
\bibitem{Anderson:2017aux} 
  L.~B.~Anderson, X.~Gao, J.~Gray and S.~J.~Lee,
  ``Fibrations in CICY Threefolds,''
  JHEP {\bf 1710}, 077 (2017)
  doi:10.1007/JHEP10(2017)077
  [arXiv:1708.07907 [hep-th]].
  
\bibitem{Morrison:2016lix} 
  D.~R.~Morrison, D.~S.~Park and W.~Taylor,
  ``Non-Higgsable abelian gauge symmetry and F-theory on fiber products of rational elliptic surfaces,''
  arXiv:1610.06929 [hep-th].

\bibitem{Morrison:2014era} 
  D.~R.~Morrison and W.~Taylor,
  ``Sections, multisections, and U(1) fields in F-theory,''
  arXiv:1404.1527 [hep-th].
  
\bibitem{Anderson:2014yva} 
  L.~B.~Anderson, I.~García-Etxebarria, T.~W.~Grimm and J.~Keitel,
  ``Physics of F-theory compactifications without section,''
  JHEP {\bf 1412}, 156 (2014)
  doi:10.1007/JHEP12(2014)156
  [arXiv:1406.5180 [hep-th]].

 \bibitem{nakayama}
N.~Nakayama,
``On Weierstrass Models",
Alg. Geom and Comm. Alg. (1987), 405.
  
  
\bibitem{Lynker:1998pb}
  M.~Lynker, R.~Schimmrigk and A.~Wisskirchen,
  ``Landau-Ginzburg vacua of string, M theory and F theory at c = 12,''
  \textsf{\doiref{10.1016/S0550-3213(99)00204-7}{Nucl.\ Phys.\ B {\bf 550} (1999) 123}, \arxivref{hep-th/9812195}}.

\bibitem{Brunner:1996bu}
  I.~Brunner, M.~Lynker and R.~Schimmrigk,
  ``Unification of M theory and F theory Calabi-Yau fourfold vacua,''
  \textsf{\doiref{10.1016/S0550-3213(97)89481-3}{Nucl.\ Phys.\ B {\bf 498} (1997) 156}, \arxivref{hep-th/9610195}}.
  
\bibitem{cicylist4} The full list of CICY four-fold configuration matrices and their Euler characteristics can be downloaded at \href{http://www-thphys.physics.ox.ac.uk/projects/CalabiYau/Cicy4folds/index.html}{{\tt http://www-thphys.physics.ox.ac.uk/projects/CalabiYau/Cicy4folds/index.html}}.


  
\bibitem{Kreuzer:2000xy} 
  M.~Kreuzer and H.~Skarke,
  ``Complete classification of reflexive polyhedra in four-dimensions,''
  Adv.\ Theor.\ Math.\ Phys.\  {\bf 4}, 1209 (2002)
  doi:10.4310/ATMP.2000.v4.n6.a2
  [hep-th/0002240].

  
\bibitem{Braun:2010vc}
  V.~Braun,
  ``On Free Quotients of Complete Intersection Calabi-Yau Manifolds,''
  \textsf{\doiref{10.1007/JHEP04(2011)005}{JHEP {\bf 1104}, 005 (2011)}, \arxivref{1003.3235}}.

\bibitem{Candelas:1987kf}
  P.~Candelas, A.~M.~Dale, C.~A.~Lutken and R.~Schimmrigk,
  ``Complete Intersection Calabi-Yau Manifolds,''
  \textsf{\doiref{10.1016/0550-3213(88)90352-5}{Nucl.\ Phys.\ B {\bf 298} (1988) 493}}.


\bibitem{Anderson:2018heq} 
  L.~B.~Anderson, A.~Grassi, J.~Gray and P.~K.~Oehlmann,
  ``F-theory on Quotient Threefolds with (2,0) Discrete Superconformal Matter,''
  arXiv:1801.08658 [hep-th].
  


\bibitem{Hubsch:1986ny}
  T.~Hubsch,
  ``Calabi-Yau Manifolds: Motivations And Constructions,''
  \textsf{\doiref{10.1007/BF01210616}{Commun.\ Math.\ Phys.\  {\bf 108} (1987) 291}}.


\bibitem{Green:1986ck}
  P.~Green and T.~Hubsch,
  ``Calabi-Yau Manifolds As Complete Intersections In Products Of Complex Projective Spaces,''
  \textsf{\doiref{10.1007/BF01205673}{Commun.\ Math.\ Phys.\ {\bf 109} (1987) 99}}.
  
\bibitem{Candelas:1987du}
  P.~Candelas, C.~A.~Lutken and R.~Schimmrigk,
  ``Complete Intersection Calabi-Yau Manifolds. 2. Three Generation Manifolds,''
  \textsf{\doiref{10.1016/0550-3213(88)90173-3}{Nucl.\ Phys.\ B {\bf 306} (1988) 113}}.
  
\bibitem{Hubsch:1992nu}
  T.~Hubsch,
  ``Calabi-Yau manifolds: A Bestiary for physicists,''
  Singapore, Singapore: World Scientific (1992) 362 p.


  \bibitem{Anderson:2015iia} 
  L.~B.~Anderson, F.~Apruzzi, X.~Gao, J.~Gray and S.~J.~Lee,
  ``A new construction of Calabi-Yau manifolds: Generalized CICYs,''
  Nucl.\ Phys.\ B {\bf 906}, 441 (2016)
  doi:10.1016/j.nuclphysb.2016.03.016
  [arXiv:1507.03235 [hep-th]].


  \bibitem{Gray:2013mja} 
  J.~Gray, A.~S.~Haupt and A.~Lukas,
  ``All Complete Intersection Calabi-Yau Four-Folds,''
  JHEP {\bf 1307}, 070 (2013)
  doi:10.1007/JHEP07(2013)070
  [arXiv:1303.1832 [hep-th]].


  
  
\bibitem{Candelas:2008wb}
P.~Candelas and R.~Davies, ``{New Calabi-Yau Manifolds with Small Hodge
  Numbers},'' {\em Fortsch.Phys.} {\bf 58} (2010) 383--466,
\href{http://arXiv.org/abs/0809.4681}{{\tt 0809.4681}}.

\bibitem{Candelas:2010ve}
P.~Candelas and A.~Constantin, ``{Completing the Web of $Z_3$ - Quotients of
  Complete Intersection Calabi-Yau Manifolds},'' {\em Fortsch.Phys.} {\bf 60}
  (2012) 345--369,
\href{http://arXiv.org/abs/1010.1878}{{\tt 1010.1878}}.

\bibitem{Candelas:2015amz}
P.~Candelas, A.~Constantin, and C.~Mishra, ``{Hodge Numbers for CICYs with
  Symmetries of Order Divisible by 4},'' {\em Fortsch. Phys.} {\bf 64} (2016),
  no.~6-7, 463--509,
\href{http://arXiv.org/abs/1511.01103}{{\tt 1511.01103}}.

\bibitem{Candelas:2016fdy}
P.~Candelas, A.~Constantin, and C.~Mishra, ``{Calabi-Yau Threefolds With Small
  Hodge Numbers},''
\href{http://arXiv.org/abs/1602.06303}{{\tt 1602.06303}}.

\bibitem{Constantin:2016xlj} 
  A.~Constantin, J.~Gray and A.~Lukas,
  ``Hodge Numbers for All CICY Quotients,''
  JHEP {\bf 1701}, 001 (2017)
  doi:10.1007/JHEP01(2017)001
  [arXiv:1607.01830 [hep-th]].



\bibitem{cicylist} The original list of CICY three-folds found in ref.~\cite{Candelas:1987kf} can be downloaded at 
\href{http://www-thphys.physics.ox.ac.uk/projects/CalabiYau/cicylist/index.html}{{\tt http://www-thphys.physics.ox.ac.uk/projects/CalabiYau/cicylist/index.html}} and, in its original format, at 
\href{http://www.th.physik.uni-bonn.de/th/People/netah/cy/cicys/cicy.html}{{\tt http://www.th.physik.uni-bonn.de/th/People/netah/cy/cicys/cicy.html}}.

\bibitem{Anderson:2009mh} 
  L.~B.~Anderson, J.~Gray, Y.~H.~He and A.~Lukas,
  ``Exploring Positive Monad Bundles And A New Heterotic Standard Model,''
  JHEP {\bf 1002}, 054 (2010)
  doi:10.1007/JHEP02(2010)054
  [arXiv:0911.1569 [hep-th]].

\bibitem{Donagi:2004ub} 
  R.~Donagi, Y.~H.~He, B.~A.~Ovrut and R.~Reinbacher,
  ``The Spectra of heterotic standard model vacua,''
  JHEP {\bf 0506}, 070 (2005)
  doi:10.1088/1126-6708/2005/06/070
  [hep-th/0411156].

  
  \bibitem{Anderson:2015yzz} 
  L.~B.~Anderson, F.~Apruzzi, X.~Gao, J.~Gray and S.~J.~Lee,
  ``Instanton superpotentials, Calabi-Yau geometry, and fibrations,''
  Phys.\ Rev.\ D {\bf 93}, no. 8, 086001 (2016)
  doi:10.1103/PhysRevD.93.086001
  [arXiv:1511.05188 [hep-th]].
  
  \bibitem{Anderson:2016ler} 
  L.~B.~Anderson, X.~Gao, J.~Gray and S.~J.~Lee,
  ``Tools for CICYs in F-theory,''
  JHEP {\bf 1611}, 004 (2016)
  doi:10.1007/JHEP11(2016)004
  [arXiv:1608.07554 [hep-th]].
      
  \bibitem{kollar-criteria}
J.~Koll\'ar,
``Deformations of elliptic Calabi--Yau manifolds,''
[arXiv:1206.5721 [math.AG]].

\bibitem{ogu}
Keiji Oguiso, 
``On algebraic fiber space structures on a Calabi-Yau 3-fold,'' 
Internat. J. Math. 4 (1993), no. 3, 439-465, With an appendix by Noboru Nakayama. MR 1228584
(94g:14019).

\bibitem{wil}
P.~M.~H.~Wilson, 
``The existence of elliptic fibre space structures on Calabi-Yau threefolds,'' 
Math. Ann. 300 (1994), no. 4, 693-703. MR 1314743 (96a:14047). 


\bibitem{Bhardwaj:2015oru} 
  L.~Bhardwaj, M.~Del Zotto, J.~J.~Heckman, D.~R.~Morrison, T.~Rudelius and C.~Vafa,
  ``F-theory and the Classification of Little Strings,''
  Phys.\ Rev.\ D {\bf 93}, no. 8, 086002 (2016)
  doi:10.1103/PhysRevD.93.086002
  [arXiv:1511.05565 [hep-th]].
  
\bibitem{deBoer:2001wca} 
  J.~de Boer, R.~Dijkgraaf, K.~Hori, A.~Keurentjes, J.~Morgan, D.~R.~Morrison and S.~Sethi,
  ``Triples, fluxes, and strings,''
  Adv.\ Theor.\ Math.\ Phys.\  {\bf 4}, 995 (2002)
  doi:10.4310/ATMP.2000.v4.n5.a1
  [hep-th/0103170].
  
\bibitem{thefiblist}
The standard list of CICY three-fold fibrations, first produced in \cite{Anderson:2017aux} can be found here: http://www1.phys.vt.edu/cicydata/.

 
  \bibitem{schoen}
 C.~Schoen, ``On Fiber Products of Rational Elliptic Surfaces with Section", Mathematische Zeitschrift 197.2 (1988): 177-200.
 
  
\bibitem{Braun:2017juz} 
  A.~Braun, A.~Lukas and C.~Sun,
  ``Discrete Symmetries of Calabi-Yau Hypersurfaces in Toric Four-Folds,''
  C. Commun. Math. Phys. (2017)
  doi:10.1007/s00220-017-3052-1
  [arXiv:1704.07812 [hep-th]].


\bibitem{Hori:2006dk} 
  K.~Hori and D.~Tong,
  ``Aspects of Non-Abelian Gauge Dynamics in Two-Dimensional N=(2,2) Theories,''
  JHEP {\bf 0705}, 079 (2007)
  doi:10.1088/1126-6708/2007/05/079
  [hep-th/0609032].
  
\bibitem{Jockers:2012zr} 
  H.~Jockers, V.~Kumar, J.~M.~Lapan, D.~R.~Morrison and M.~Romo,
  ``Nonabelian 2D Gauge Theories for Determinantal Calabi-Yau Varieties,''
  JHEP {\bf 1211}, 166 (2012)
  doi:10.1007/JHEP11(2012)166
  [arXiv:1205.3192 [hep-th]].
  
\bibitem{Hori:2013gga} 
  K.~Hori and J.~Knapp,
  ``Linear sigma models with strongly coupled phases - one parameter models,''
  JHEP {\bf 1311}, 070 (2013)
  doi:10.1007/JHEP11(2013)070
  [arXiv:1308.6265 [hep-th]].
  
\bibitem{Caldararu:2017usq} 
  A.~Caldararu, J.~Knapp and E.~Sharpe,
  ``GLSM realizations of maps and intersections of Grassmannians and Pfaffians,''
  JHEP {\bf 1804}, 119 (2018)
  doi:10.1007/JHEP04(2018)119
  [arXiv:1711.00047 [hep-th]].

  \bibitem{gross_mult}
  M. Gross, ``Elliptic Three-folds II: Multiple Fibers", 1997. Trans.Am.Math.Soc.,349,3409-3468


   
\bibitem{Cecotti:2010bp} 
  S.~Cecotti, C.~Cordova, J.~J.~Heckman and C.~Vafa,
  ``T-Branes and Monodromy,''
  JHEP {\bf 1107}, 030 (2011)
  doi:10.1007/JHEP07(2011)030
  [arXiv:1010.5780 [hep-th]].
  
\bibitem{Donagi:2011jy} 
  R.~Donagi and M.~Wijnholt,
  ``Gluing Branes, I,''
  JHEP {\bf 1305}, 068 (2013)
  doi:10.1007/JHEP05(2013)068
  [arXiv:1104.2610 [hep-th]].
  
\bibitem{Anderson:2017zfm} 
  L.~B.~Anderson, L.~Fredrickson, M.~Esole and L.~P.~Schaposnik,
  ``Singular Geometry and Higgs Bundles in String Theory,''
  SIGMA {\bf 14}, 037 (2018)
  doi:10.3842/SIGMA.2018.037
  [arXiv:1710.08453 [math.DG]].
  
\bibitem{Anderson:2017rpr} 
  L.~B.~Anderson, J.~J.~Heckman, S.~Katz and L.~Schaposnik,
  ``T-Branes at the Limits of Geometry,''
  JHEP {\bf 1710}, 058 (2017)
  doi:10.1007/JHEP10(2017)058
  [arXiv:1702.06137 [hep-th]].
  
\bibitem{Anderson:2013rka} 
  L.~B.~Anderson, J.~J.~Heckman and S.~Katz,
  ``T-Branes and Geometry,''
  JHEP {\bf 1405}, 080 (2014)
  doi:10.1007/JHEP05(2014)080
  [arXiv:1310.1931 [hep-th]].
  
   
 \end{thebibliography}
\end{document}